# Titan at the time of the Cassini spacecraft first flyby: a prediction for its origin, bulk chemical composition and internal physical structure

**A. J. R. Prentice** *

School of Mathematical Sciences

Monash University, Victoria 3800, AUSTRALIA

* E-mail: andrew.prentice@sci.monash.edu.au

Address for Correspondence :  School of Mathematical Sciences
Monash University
Victoria  3800
AUSTRALIA

## ABSTRACT

I report the results of a new set of calculations for the gravitational contraction of the proto-solar cloud to quantify the idea that Titan may be a captured moon of Saturn (Prentice 1981, 1984). It is proposed that Titan initially condensed as a secondary embryo in the same protosolar gas ring from which Saturn's central solid core and gaseous envelope were acquired. The new model for the protosolar cloud (hereafter PSC) includes the influence of very strong superadiabaticity in its outer layers that is observed in an exact numerical simulation of supersonic turbulent convection (Prentice & Dyt 2003). I adopt the protosolar elemental abundances recommended by Lodders (2003).

At Saturn's orbit, the bulk chemical constituents of the condensate are rock (mass fraction 0.494), water ice (0.474), and graphite (0.032). The mean density is 1523 kg m$^{-3}$. Structural models for a frozen Titan yield a mean density of 2095 kg m$^{-3}$ (chemically homogeneous case) and 1904 kg m$^{-3}$ (fully differentiated 2-zone case). The agreement to one percent of the latter value with the observed mean density suggests that Titan is indeed a fully differentiated satellite. The value of $C/MR^2$ for this model is 0.316. It is predicted that Titan has no internal ocean or induced magnetic field but it may possess a small native dipole field of magnitude $2 \times 10^{11}$ Tesla m$^3$ due to thermoremanent magnetization fed by Saturn's ancient magnetic field. Capture of Titan was achieved by gas drag at the edge of the proto-Saturnian envelope at a time when that cloud had a radius close to Titan's present orbital size. Collisional drag was also probably an important agent in securing Titan's capture. Perhaps Hyperion is the shattered remnant of a pre-existing native moon of Saturn that was destroyed on Titan's arrival. Titan should thus have much the same appearance as Triton, being nearly smooth, crater-free and streaked with elemental carbon (Prentice 2004a).

# 1. INTRODUCTION

## 1.1 Background

The eagerly awaited release of scientific data from the first close encounter of the Cassini-Huygens spacecraft with Titan on 2004 October 26 has generated fresh interest in the origin and internal structure of this unusual satellite of Saturn (Rappaport et al. 1997; Matson, Spilker & Lebreton 2002; Lunine & Soderblom 2002). Aside from the fact that Titan is the only natural satellite in the Solar system to possess an atmosphere so dense and opaque that the satellite surface is hidden from normal view, the thing that makes Titan unusual is its sheer size. Its physical radius $R_{Ti} = 2575 \pm 2$ km and mean density of $1881 \pm 5$ kg m$^{-3}$ mean that Titan is nearly as big and massive as Jupiter's largest satellite Ganymede, even though Saturn's mass is only ~0.3 times that of Jupiter. Second, Titan is ~58 times more massive than Saturn's next largest moon Rhea. As shown below, such disparities are very difficult to reconcile if Titan formed from Saturn's primitive circum-planetary cloud by the same physical process by which it is proposed that Saturn's other intermediate-sized regular satellites had formed, and by which process the system of Galilean satellites had also formed from Jupiter's primitive cloud.

To understand Titan's origin Prentice (1981, 1984) has suggested that Titan is not a native moon of Saturn. Instead, it originally condensed as a secondary planetary embryo within the gas ring that was shed by the protosolar cloud (PSC) at Saturn's heliocentric distance, and was later captured by Saturn. The shedding of a concentric system of orbiting gas rings is a central feature of the author's modern Laplacian theory (hereafter MLT) of Solar system origin (Prentice 1978a,b 2001). In this model, Saturn's rock/ice core of mass ~15–20$M_\oplus$ (Anderson et al. 1980) was the primary solid embryo within that protosolar gas ring. This core then accreted a large quantity of gas and residual solids from the gas ring to form Saturn's extended proto-planetary cloud (cf. Mizuno 1980). Next, this primitive cloud, whose initial size may have been ~$60R_S$ ($R_S$ = 60,268 km = Saturn's equatorial radius) proceeded to dispose of its excess spin angular momentum through the detachment of its own concentric family of gas rings. The major icy satellites of Saturn, including Rhea, condensed from this system of rings.

It is now supposed that Titan entered the Saturn system on a prograde orbit at a time when the radius of the cloud was close to the present orbital radius of Titan, namely ~$20.27R_S$. Aerodynamic and collisional drags were the most likely mechanisms for capture (Pollack, Burns & Tauber 1979). Neither Rhea nor any of the satellites interior to that orbit ($a_{Rhea}$ = $8.7R_S$) would have been formed at that stage. It is conceivable, however, that Hyperion is a shattered remnant of a pre-existing regular satellite of Saturn that had formed at orbital radius ~20–$25R_S$ prior to Titan's arrival, and was then struck by Titan. Perhaps Iapetus was displaced outwards to its present orbital distance, namely $a_{Iapetus}$ = $59R_S$ by the capture process.

It is the purpose of this paper to report detailed predictions for bulk chemical composition and internal physical structure of Titan, based on the capture origin hypothesis. These predictions, which are shortly to be tested by the Cassini-Huygens spacecraft, will provide a valuable test of the MLT as a viable theory of solar system origin.

## 2   THE MODERN LAPLACIAN THEORY

### 2.1   Contraction of the protosolar cloud and the shedding of gas rings

According to the MLT, both the planetary system and the regular satellite systems of the major planets accreted close to their present orbital locations from circular rings of gas that had been cast off by the PSC and the proto-planetary clouds of the gas giant planets, respectively. The gas rings are shed from the equatorial region of the parent cloud as a means of disposing of excess spin angular momentum during gravitational contraction. In the case of the PSC, it is possible that the large Kuiper Belt object Quaoar marks the orbit of the very first ring of gas shed by that cloud. Quaoar moves on a resonance-free orbit that has low eccentricity and modest inclination (Trujillo & Brown 2003). The ratio of its present mean heliocentric distance $R_0 = 9{,}323 R_\odot$ ($R_\odot = 6.9598 \times 10^5$ km = Solar radius) to that of Neptune ($R_1$) is $R_0/R_1 = 1.442$. This number is consistent with the other neighbouring planet distance ratios $R_n/R_{n+1}$ ($n = 1,2,3,...$) in the Solar system. The mean ratio of $R_n/R_{n+1}$ from Mercury through to Quaoar is $1.70 \pm 0.22$.

Let $M_n$ and $f_n$ denote the residual mass and moment-of-inertia factor of a uniformly rotating gas cloud at the moment of detachment $t_n$ of a gas ring of mass $m_n$ at orbital distance $R_n$. If $f_n \ll 1$, the laws of conservation of total mass and angular momentum yield

$$R_n/R_{n+1} \approx \left(1 + \frac{m_{n+1}}{M_{n+1} f_{n+1}}\right)^2 \qquad (1)$$

If the cloud contracts homologously, so that both $f_n$ and $m_n/M_n$ stay sensibly constant, then so also is $R_n/R_{n+1}$ a constant. That is, the radii $R_n$ ($n = 0, 1, 2, 3,...$) form a geometric sequence. This accounts for the basic geometric structure of the progression of planetary distances and of the orbital distance sequences of the regular satellite systems. If we apply equation (1) to the PSC and set $M_n \approx M_\odot$, $f_n \approx 0.01$, we find $m_n \approx 0.003 M_\odot \approx 6 \times 10^{27}$ kg.

Now, adopting the protosolar elemental abundances recommended by Lodders (2003) and sequestering all rock-like elements into oxides and sulphides and the ice-like elements (C, N, O) into hydrides (CH$_4$, NH$_3$, H$_2$O), the total mass of condensable rock and ice in each ring is $m_{cond} \approx 15 M_\oplus$. This tallies well with the observed masses of Uranus and Neptune, and with the observed core mass of Saturn noted earlier.

**2.2　Application of the mass-distance relations to the proto-Saturnian cloud**

We now apply equation (1) to the inner Saturnian satellite system and set $\langle R_n / R_{n+1} \rangle = 1.300$, which is the observed mean orbital distance ratio taken from Mimas through to Rhea. Taking $f_n = 0.01$ as before and $M_n = M_S$ (= Saturn's mass), we obtain a mean gas ring mass $m_n = 0.0014 M_S = 8.0 \times 10^{23}$ kg. Again adopting protosolar elemental abundances, the mass of rock + ice (H$_2$O & NH$_3$) condensate in each gas ring is $m_{cond} = 9.4 \times 10^{21}$ kg. The mass of Rhea ($2.32 \times 10^{21}$ kg) is consistent with $m_{cond}$ if we allow for inefficiency in the process of delivery of condensate grains onto the mean orbit of the gas ring, where gravitational accumulation of the satellite takes place. We note here that a reduced efficiency of satellite accumulation is expected in the Saturnian system owing to the much closer orbital spacings of the gas rings. Diffusive mixing of gas at the cylindrical boundary surfaces between adjacent rings destroys the unique orbital angular velocity distribution of the gas which is responsible for focusing the condensate grains onto the mean orbit $R_n$ (Prentice 1978a).

## 2.3 The case against Titan as a native satellite of Saturn

The important fact which now emerges from the above discussion is that Titan's mass, namely $M_{Ti} = 1.3455 \times 10^{23}$ kg, exceeds $m_{cond}$ by a factor of ~14. If the MLT is worth its salt, Titan cannot be a native satellite of Saturn.

# 3 SUPERSONIC TURBULENT CONVECTION AS THE DRIVING MECHANISM OF THE MLT

## 3.1 The need for large-scale turbulent stress and viscosity

In order for the contracting PSC to shed its excess spin angular momentum in a series of isolated gas rings, it is necessary (i) that the interior of the cloud rotate nearly uniformly like a solid body, (ii) that the moment-of-inertia factor of the cloud be extremely small, typically $f_n \approx 0.01 - 0.02$, and (iii) that there be a very steep density inversion at the surface of the cloud, with the mean density $\rho$ rising by a factor of 30–40. Previously it has been demonstrated that the above 3 circumstances can be achieved if the interior of the PSC is pervaded by powerful buoyancy-driven convective currents whose speeds $v_t$ just below the photosurface greatly exceed the local adiabatic sound speed $v_s = \sqrt{\gamma p_{gas}/\rho}$. Here $p_{gas} = \rho \Re T / \mu$ is the gas pressure, $\Re$ is the gas constant, $T$ the temperature, $\mu$ is the mean molecular weight and $\gamma$ is the ratio of specific heats (Prentice 1973, 1978a, 2001).

Very strong thermal convection creates a radial turbulent stress $p_{turb} = \langle \rho_t v_t^2 \rangle$ which alters the internal mean density profile of the cloud in the direction of lowering $f_n$. The turbulent convection also creates a turbulent viscosity $\eta_{turb}$ which serves to eliminate spatial gradients in the angular velocity distribution. Lastly, if all radial motions cease at the photosurface, a very steep density inversion arises naturally from the degradation of turbulent stress. If $\rho_-$ denotes the gas density of fully turbulent material just below the photosurface and $\rho_+$ the density at the base of the lid of non-turbulent material just above it, we have

$$\rho_+ / \rho_- = 1 + \left[ \langle \rho_t v_t^2 \rangle / p_{\text{gas}} \right]_- . \qquad (2)$$

### 3.2 Difficulties with the previous non-superadiabatic models

If rotation is ignored, then $p_{\text{turb}} = \beta \rho G M(r)/r$, where $M(r)$ is the mass interior to radius $r$, $G$ is the gravitation constant and $\beta \approx 0.1$ is called the turbulence parameter. It is here that a physical problem with the MLT occurs. In order to achieve a density inversion factor $\rho_+ / \rho_- \approx 30 - 40$, we shall require $\gamma \langle \rho_t v_t^2 \rangle / \rho v_s^2 \approx 35$. This is considered unlikely since it implies convective speeds $v_t \geq 5 v_s$, taking $\gamma = 1.4$. Detailed numerical simulations of very strong convection in thermally unstable atmospheres that are stratified gravitationally across many pressure scale heights indicate that the maximum Mach number $M = \langle v_t \rangle / v_s$ is always less than ~3 (Cattaneo, Hurlburt & Toomre 1990). This means that the ratio $F_t \equiv \langle \rho_t v_t^2 \rangle / p_{\text{gas}} \leq 12$.

### 3.3 The role of strong superadiabaticity in achieving a physical model of supersonic turbulent convection

To resolve the above difficulties, Prentice & Dyt (2003) conducted a numerical simulation of supersonic turbulent convection in a model atmosphere to mimic the physical conditions in the upper layer of the PSC. A sub-grid scale turbulence approximation (Smagorinsky 1963) was used to model the influence of motions whose length scale is less than the computational grid-size. Adopting the same boundary conditions as those used by Cattaneo et al. (1990), Prentice & Dyt discovered that the decline of all vertical motion near the top boundary (corresponding to depth $z = 0$) of the atmosphere caused that region to become strongly superadiabatic and to acquire a negative polytropic index $n_t = \partial(\log_e \rho)/\partial(\log_e T)$. A natural upturn in the horizontally-averaged density $\langle \rho(z) \rangle_h$ then accompanies a sharp steepening of the average vertical temperature gradient as $z \to 0$. Peak vertical convective speeds $v_{t,\text{peak}}$ are achieved by spatially-concentrated downflows in the deep atmosphere where the surrounding gas is essentially adiabatic, having $\langle n_t \rangle_h \approx 2.5$.

In a typical simulation where $v_{t,\text{peak}} \approx 1.1 v_s$, $\langle \rho(0) \rangle_h$ rises to 3.5 times the density $\rho_0$ at the top boundary of the initial quiescent atmosphere. At the same time $\langle n_t \rangle_h$ falls to –0.65 at $z = 0$. Since $\langle \rho(0) \rangle_h \propto (v_{t,\text{peak}})^2$, it is now feasible that the inversion factor $\langle \rho(0) \rangle_h / \rho_0 \approx 30$ needed to validate the MLT may be achieved for numerical simulations having peak Mach number $M \approx 3.2$. This is almost physically acceptable.

## 4. A NEW PROTOSOLAR CLOUD

### 4.1 Structural equations of the non-rotating cloud

A new numerical model for the structure and gravitational contraction of the PSC, which incorporates the above findings, has been constructed. The first step is the determination of the internal structure of a non-rotating spherically-symmetric cloud of given surface radius $R_s$ and mass $M_s$. If $T(r)$ denotes the temperature at radius $r$, and $T_c = T(0)$ is the central temperature, we define a dimensionless temperature function $\theta(r) = \mu_c T(r)/\mu T_c$. Here $\mu = \mu(r)$ is the mean molecular weight and $c$ refers to the centre. The cloud elemental abundances are the protosolar values recommended by Lodders (2003), with the one minor exception of calcium. For this element the abundance of Ca relative to Fe has been set equal to the CI chondritic ratio, namely $[n(Ca)/n(Fe)]_{CI} = 0.069$, rather than the photospheric value. This change is extremely important because the planet Mercury, which is thought to be rich in calcium (Lewis 1972), provides an extremely valuable calibration point in determining the bulk chemical composition of the planetary system (see below). CI chondrites are highly representative of the elemental abundances of the inner Solar system condensate.

Each PSC model of given radius $R_s$ consists of an inner adiabatic convective core of radius $r_1$ in which the radial turbulent stress is given by $p_{turb} = \beta_1 \rho GM(r)/r$, where $\beta_1$ is a constant parameter. Next, there is a 2$^{nd}$ radius point $r_2$ inside which all hydrogen is assumed to be H (or H$^+$). Beyond $r_2$ all hydrogen is taken to be molecular H$_2$. The adiabatic core is surrounded by a superadiabatic envelope of polytropic index $n_{sup} = -1$ in which the turbulence parameter $\beta = \beta(r)$ declines linearly with the temperature difference $T(r) - T_s$ according as

$$\beta(r) = \beta_1 \cdot (\theta(r) - \theta_s)/(\theta_1 - \theta_s),$$

where $\theta_1 = \mu_c T(r_1)/\mu_1 T_c$. Clearly $\beta(r_1) = \beta_1$.

### 4.2 Inclusion of rotation and the specification of the model controlling parameters

The next step in the construction of the PSC model is to include rotation. This is done using the atmospheric approximation (Jean 1928, Prentice 1978a). The radius $R_s$ now corresponds to the polar radius $R_p$ of the rotating cloud and if the cloud is fully rotating, meaning that the centripetal acceleration at the equator matches inward gravitational acceleration, the equatorial radius $R_e$ is given $R_e = \frac{3}{2}R_s$. The controlling parameter of a cloud of given mass $M_s$ and polar radius $R_s$ are then $\beta_1$, $\theta_1$ and $\theta_s$. If these numbers are kept constant during the contraction, the PSC sheds a system of gas rings whose mean orbital radii $R_n$ ($n = 0,1,2,...$) form a nearly geometric sequence. The parameter $\theta_s$ is chosen so that the observed present distance of Jupiter relative to Mercury is accounted for after allowing for the secular expansion of orbits due to mass loss from the PSC. If $R_{n,i}$ and $M_{n,i}$ denote the mass of the PSC at the time $t_n$ of detachment of the $n^{th}$ ring, the present orbital radius of the planetary condensate is given simply by $R_n = R_{n,i} M_n/M_{Sun}$.

The parameter $\beta_1$ is adjusted so that the density of the condensate at Mercury's orbit has a bulk mean density which coincides with the observed mean uncompressed density of the planet, namely 5300 kg m$^{-3}$ (Anderson et al. 1987; Taylor 1991). The parameter $\theta_1$ determines the depth of the superadiabatic outer envelope of the PSC. This parameter also has an influence on the bulk chemical composition of the condensate that forms at Mercury's orbit, as well as that of the condensate at Jupiter's orbit, as follows. First, increasing the gas ring mean orbit pressure $p_n$ increases the separation between the condensation temperatures of Fe–Ni metal and the principal rock species $MgSiO_3$–$Mg_2SiO_4$ in the Mercurian gas ring. This enhances the metal fraction of the solid there (Lewis 1972, Prentice 1991). Second, increasing $p_n$ also elevates the condensation temperature of water ice in the gas ring from which Jupiter formed. With $\beta_1 = 0.12442$ chosen to account for the mean density of Mercury, the choices $\theta_1/\theta_s = 8.2$ and $\theta_s = 0.002250$ simultaneously ensure not only the correct final position of Mercury relative to Jupiter, but also that only a mass fraction $\phi_{H_2O} = 0.665$ of the total water vapour at Jupiter's orbit can condense out. This calibration marker is important since it means that the condensate that is later swept up to form the heavy element compound of Jupiter's atmosphere is enhanced in rock relative to $H_2O$ by just the right proportion to explain the super-solar rock-to-ice mass fractions of Ganymede, namely 0.55:0.45 (Prentice 1996a, 2001).

### 4.3 Initial conditions and carbon chemistry in the PSC

The PSC is assumed to acquire quasi-hydrostatic and thermodynamic equilibrium at initial polar radius $8000R_\odot$. It is assumed that the first gas ring is cast off at the orbit of Quaoar, where the polar radius of the cloud is $6215R_\odot$. The mass of this gas ring is $0.00121M_\odot$ and the residual mass of the cloud then is $1.19565M_\odot$. Initially, all carbon at the photosurface of the PSC is in the form of CO. A steady conversion of CO to both $CH_4$ and pure graphite C(s) then takes place through catalytic synthesis on the surfaces of small Fe–Ni grains deep within the atmosphere, as explained elsewhere (Prentice 1993, 2001). Rapid inter-conversion of CO to $CO_2$ also takes place through the water-shift gas phase reaction. Selecting a nominal synthesis temperature $T_{cat} = 462.4\,\text{K}$ and mean Fe–Ni grain radius $a_{Fe-Ni} = 5\times10^{-7}\,\text{m}$, the production of pure carbon achieves a maximum at cloud equatorial radius $R_e = 2000R_\odot$ and all C(s) disappears for $R_e < 750R_\odot$. This radius lies mid-way between the initial orbits of Jupiter and the main belt asteroids.

# 5  GRAVITATIONAL CONTRACTION OF THE PROTOSOLAR CLOUD.

## 5.1  Carbon chemistry and chemical condensation in the outer system of protosolar gas rings

Table 1 gives the residual mass $M_n$ of the protosolar cloud and the time of detachment $t_n$ of the system of protosolar gas rings whose initial mean orbital radii $R_{n,i}$ lead to planets at the present-day orbital radii $R_n$ ($n = 0,1,2,...$) of Quaoar, Neptune, Uranus, Saturn and Jupiter, respectively. As mentioned above in Section 4, the cloud is assumed to establish quasi-hydrostatic and thermodynamic equilibrium at initial polar radius $8000 R_\odot$. This choice ensures (see Section 5.2) that the mass fraction of $CH_4$ ice in Phoebe is just the right amount to yield a mean density for this moon which matches the 2004 June 11 Cassini measurement.

The last four columns of Table 1 show the distribution of carbon in its four principal compound states for each protosolar gas ring.  The main point to note in this table is the steady conversion of CO and $CO_2$ to $CH_4$ and C(s) that takes place during the contraction of the PSC.  The other point to note is that the production of graphite C(s) is only a transient phenomenon.  By the time that the cloud has shrunk to the orbit of the main belt asteroids, where $R_{5,i} = 528 R_\odot$, all C(s) has disappeared along with CO and $CO_2$.

Table 2 lists the physical properties of the outer family of gas rings and the expected bulk chemical composition of the condensate grains which settle onto the mean orbit of each ring. Equilibrium condensation chemistry is assumed along the lines first advanced by Lorimer (1967), Lewis (1972) and Grossman (1972).  The computational code which determines the bulk composition of the condensate been specifically designed to deal with the gas ring configuration of the MLT (Prentice 1991, 1996a, 2001). The final column of the table gives the mean density of the condensate. This is calculated for the blackbody temperature $T_{b,n}$ at the present-day orbital distances $R_n$. It is given by $T_{b,n} = 3430 \, (R_\odot / R_n)^{0.5}$ K for an assumed uniform albedo of 0.5.

## 5.2 Phoebe's origin as a captured Kuiper Belt object

The flyby of Phoebe by the Cassini spacecraft on 2004 June 11 revealed this small moon of Saturn to be an ancient, heavily-cratered ice ball. Phoebe has a mean radius of 107 km and a mean density of 1600 kg m$^{-3}$. The Cassini VIMS experiment did show that the surface consists of materials similar to those seen on Pluto and Triton. Widespread deposits of $CO_2$ were detected, along with water ice and other carbon-bearing compounds (see NASA news release 2004-158, dated 2004 June 23).

The unusual retrograde orbit of Phoebe about Saturn suggests to most scientists that Phoebe is a captured body. Prior to the Cassini flyby, Prentice (2004b) predicted that Phoebe had originally condensed at Saturn's orbit and, like Titan, would consist solely of rock, water ice and graphite. He also predicted a Phoebe mean density of 1330 kg m$^{-3}$ based on pre-Lodders (2003) solar elemental abundances. These earlier abundances lead to a condensate that is much icier – and hence of lower mean density - owing to the higher O abundance.

Inspecting Table 2, it is clear that an origin for Phoebe as a remnant of the gas ring from which Quaoar and the other Kuiper Belt objects formed is the one which best fits the Cassini data. Although the new predicted condensate density at Saturn's orbit, based on the Lodders' (2003) abundances, is not inconsistent with the Phoebe mean density, this model cannot easily explain the ubiquitous presence of $CO_2$ on Phoebe's surface. $CO_2$ ice has long been predicted by the MLT to be a major constituent of outer Solar system bodies, such as Triton and Pluto (Prentice 1990, 1993).

# 6     RESULTS: STRUCTURAL MODELS FOR TITAN

## 6.1     Discussion of the computational codes

The final section of this paper is devoted to Titan and the calculation of its physical properties, based on the notion that this large moon is a remnant condensate of the protosolar gas ring at Saturn's orbit.

Two different numerical models for Titan have been computed, one assuming the satellite is chemically uniform, and the second that it has fully differentiated into a 2-zone structure composed of a rock-graphite core and a pure water ice envelope. The first stage in the construction of each satellite model is the determination of the present-day temperature distribution. This has been achieved through use of a thermal evolutionary code that takes into account the thermophysical properties of the individual constituents of the rock. The only heating considered here is that due to the decay of the radioactive nuclides of K, U & Th. This thermal code, which was designed specifically for studying the Galilean satellites of Jupiter, is described in more detail elsewhere (Prentice 1996a, 2001). For simplicity, each model is assumed to have a uniform initial temperature equal to the present-day surface value, namely $T_{\text{surf}} = 94$ K. The ice is assumed to transfer heat by solid state convection once the diffusive temperature exceeds a fraction ~0.7 of the local melt temperature of the $H_2O$ ice (cf. Ellsworth & Schubert 1983).

A second computational code, also described in Prentice (1996a, 2001), is used to calculate the internal run of pressure and density within each satellite model, based on the temperature profile derived from by the thermal code. The code takes into account the compressibilities and thermal properties of the principal rock constituents, as well as those of the four relevant phases of $H_2O$ ice. The principal chemical constituents of the Titanian rock are $MgSiO_3$ (0.2550), $Mg_2SiO_4$ (0.1786), $SiO_2$ (0.0741), $Ca_2MgSi_2O_7$ (0.0420), $MgAl_2O_4$ (0.0320), $Fe_3O_4$ (0.1768), $FeS$ (0.1928), $NiS$ (0.0114), $Ni$ (0.0077), $NaOH$–$KOH$ (0.0123) and $NaCl$ (0.0016). The mean density of the total rock mix at 298.15 K and 0.1 Mpa is $\langle\rho\rangle_{rock} = 3646.7 \text{ kg m}^{-3}$.

### 6.2  Comparison of results for the homogeneous and 2-zone satellite modules

Figure 1 shows the internal temperature distribution $T(r,t)$ within the 2-zone model plotted as a function of radial distance $r$, which is expressed in units of the satellite radius $R_{Ti} = 2575$ km. The temperature profiles are given at several representative times $t$ during the evolution to solar age $t_{Sun} = 4600$ Myr. We notice that the temperature within the outer ice mantle increases steadily everywhere with time. The ice starts to convect first at the core boundary. This happens after about 250 Myr. The thickness of the convection layer increases to about 340 km by solar age. As the $H_2O$ mantle is totally frozen, we do not expect Titan to harbour any internal ocean. That is, no induced magnetic field is expected to be observed by the Cassini magnetometer experiment.

The broad thermophysical properties of both the chemically uniform and 2-zone differentiated models are given in Table 3. Now the observed mean density of Titan is 1881 ± 5 kg m$^{-3}$. Obviously the chemically uniform model is far too dense to be considered as a viable option for Titan. That is, we can predict immediately that Titan is a fully differentiated satellite, similar to Ganymede (Anderson et al. 1996). Bearing in mind that the abundances reported by Lodders (2003), particularly that of O, have an uncertainty of 10%, it is quite extraordinary that our first model of Titan, built up solely from first principles, has produced a model mean density that lies so close to the observed value. By reducing the ice mass fraction for Saturn in Table 2 by just ~0.0125, corresponding to a change of less than 3%, the predicted mean density of the 2-zone model coincides with the observed value.

### 6.3 Prediction of a possible native magnetic field of Titan

Lastly, consider Fig. 2 which shows the temperature at the fractional radii points of 0.375 and 0.48. These correspond to physical radii of $r_1 = 965$ and $r_2 = 1240$ km, respectively. During radiogenic evolution to present age, the temperature of the rock between these 2 points both rises above and then descends through the $Fe_3O_4$ Curie temperature. For the 0.48 radius point this event occurs at time $t = 1.5 \times 10^9$ yr. Now $Fe_3O_4$ makes up a mass fraction 0.177 of total rock. This is almost the same fraction as that of Ganymedean rock (Prentice 2001). During the course of Ganymede's radiogenic evolution to present age, the temperature at its 1470 km radius point also rose above and then fell below the Curie line at about the same time, namely $t = 1.7 \times 10^9$ yr.

It follows therefore that Titan may have experienced the same early thermal conditions as those enjoyed by Ganymede, that are the conditions necessary for the acquisition of a native magnetic dipole field through thermoremanent magnetization (Prentice 1996b, cf. Crary & Bagenal 1998). In the case of Ganymede, the source of magnetization was Jupiter's ancient magnetic field. This may have been some ~10 times stronger than the present Jovian field. The observed magnitude of the dipole moment of Ganymede is $|m|_G = 1.4 \times 10^{13}$ Tesla m$^3$ (Kivelson et al. 1998) and those of Jupiter and Saturn are $|m|_J = 1.57 \times 10^{20}$ and $|m|_S = 4.6 \times 10^{18}$ Tesla m$^3$, respectively. The strength of a planetary dipole field diminishes with equatorial distance as $1/r^3$. The orbital distances of Ganymede and Titan are $a_G = 1,070,400$ and $a_{Ti} = 1,221,900$ km. If the ratio of Saturn's primitive magnetic field to that of Jupiter at $t = 1.5 \times 10^9$ yr was the same as the present ratio, the predicted internal magnetic dipole moment of Titan is then estimated to be

$$|m|_{Ti} \approx |m|_G \cdot \left(\frac{|m|_S}{|m|_J}\right) \cdot \left(\frac{a_G}{a_{Ti}}\right)^3 \cdot \left(\frac{r_{2,Ti}}{r_{2,G}}\right)^2 \approx 0.014 |m|_G \approx 2.0 \times 10^{11} \text{ Tesla m}^3$$

This value is consistent with the upper limit of $5 \times 10^{11}$ Tesla m$^3$ to any possible internal magnetic moment found by Voyager 1 (Ness et al. 1981). It will be interesting to find what the Cassini spacecraft discovers during its first scheduled close encounter with Titan on 2004 October 26.

## 7. CONCLUSIONS

We have examined the origin of Titan within the context of the MLT. The large disparity between the mass of Titan and the expected mass of a native Saturnian satellite suggests that Titan is an interloper to the Saturnian system. We propose that Titan originally condensed as a secondary planetesimal embryo within the gas ring that was cast off by the PSC at Saturn's orbit, and that it was later captured by that planet.

A bulk chemical composition for Titan is derived as a spin-off from a new model for the gravitational contraction of the PSC which:

(i) incorporates the revised solar elemental abundances (Lodders 2003),

(ii) accounts for the observed mean spacing of the planets between the orbits of Mercury and Jupiter,

(iii) accounts for the large mean uncompressed density of Mercury,

(iv) accounts for the inferred super-solar rock-to-ice ratio (0.55:0.45) of Ganymede,

(v) accounts for the Phoebe mean density ~1600 kg m$^{-3}$ found by the Cassini spacecraft, and

(vi) provides for the deposition of pure carbon in the planetary condensate at all orbits beyond the mid-way point between the main belt asteroids and Jupiter.

The predicted bulk chemical composition of Titan is rock (mass fraction 0.494), water ice (0.074) and graphite (0.032). The mean uncompressed density of this mixture at 76 K is 1520 kg m$^{-3}$.

Two thermally-evolved structural models for Titan are then constructed on the basis of the above bulk chemical composition. It is found that the mean density of a differentiated 2-zone model, having a central rock-graphite core of mass $0.526 M_{Ti}$, exceeds the observed mean density of 1881 kg m$^{-3}$ by just 23 kg m$^{-3}$. The homogeneous model provides a very poor fit to the observed mean density constraint. Titan is thus predicted to be a fully-differentiated satellite, similar in structure to Jupiter's moon Ganymede. The mean axial moment-of-inertia coefficient of the proposed 2-zone model is $C/MR^2 = 0.316$. On the basis of the similarity between the radiogenic thermal evolutions of Titan and Ganymede, it is predicted that the Cassini spacecraft will discover that Titan has a native, internal magnetic field. The dipole moment is expected to be of order $2 \times 10^{11}$ Tesla m$^3$. No induced magnetic field is expected, consistent with a totally frozen satellite interior.

Lastly it is suggested that Titan's capture was achieved by a combination of aerodynamic and collisional braking at the edge of the proto-Saturnian cloud, when the radius of that cloud was near $20 R_S$. It is proposed that Hyperion is the remnant of a native Rhea-sized moon of Saturn that had formed within the proto-Saturnian cloud prior to Titan's arrival and was then collisionally destroyed. In any event, if Titan is a captured moon its surface must surely bear testimony of such a dynamic origin. We expect to find a surface that is similar in appearance to that of Neptune's large captured moon Triton, namely to be mostly smooth and nearly crater-free. It may also show streaks of the pure carbon which it inherited from its protosolar origin . The early blurry images of Titan that have so far been seen by Cassini are consistent with this expectation.

A short account of the results that have been described in detail here are to be presented to the 36$^{th}$ Meeting of the AAS Division for Planetary Sciences in Louisville, Kentucky, 2004 November 9 (Prentice 2004a). The electronic version of that abstract was first posted online by the AAS on 2004 September 29.

## ACKNOWLEDGMENTS

I thank John D. Anderson [NASA/JPL] and George W. Null for much generous support during numerous visits to the NASA Jet Propulsion Laboratory, Pasadena, and J. Castillo, P. D. Godfrey, R. A. Jacobson, N. J. Rappaport, E. M. Standish, E. C. Stone and K. K. Yau for valuable discussions. I am grateful to Steve Morton for much technical support.

## REFERENCES

Anderson J. D., Null G. W., Biller E. D., Wong S. K., Hubbard W. B., MacFarlane J. J., 1980, Sci, 207, 449.

Anderson J. D., Colombo G., Esposito P. B., Lau, E. L., Trager G. B., 1987, Icarus, 71, 337.

Anderson J. D., Lau, E. L., Sjogren W. L., Schubert G., Moore W. B., 1996, Nat, 384, 541.

Cattaneo F., Hurlburt N. E., Toomre J., 1990, ApJ, 349, L63.

Crary F. J., Bagenal F., 1998, J. Geophys. Res 103, 25757.

Ellsworth K., Schubert, G., 1983, Icarus, 54, 490.

Grossman L., 1972, Geochim. Cosmochim. Acta, 36, 597.

Jeans J.H., 1928, Astronomy and Cosmogony, Cambridge Univ. Press, U.K., p.389.

Kivelson M.G. et al., 1996, Nat, 384, 537.

Larimer J.W., 1967, Geochim. Cosmochim. Acta, 31, 1215.

Lewis, J.S., 1972, Earth Planet. Sci. Letts., 15, 286.

Lodders K., 2003, ApJ, 591, 1220.

Lunine, J.I., Soderblom L.A., 2002, Space Sci. Rev., 118, 191.

Matson D.L., Spilker L.J., Lebreton, J-P, 2002, Space Sci. Rev., 104, 1.

Mizuno H., 1980, Prog. Theor. Phys., 64, 544.

Ness N.H., Acuna M.H., Lepping R.D., Connerney, J.E.P., Behannon K.W., Burlaga L.F., Neubauer F.M., 1981, Sci, 212, 212.

Pollack J.B., Burns J.A., Tauber M.E., 1979, Icarus, 37, 587.

Prentice A.J.R., 1973, A&A, 27, 237

Prentice A. J. R., 1978a, Moon Planets, 19, 341.

Prentice A. J. R., 1978b, in Dermott S.F., ed., The Origin of the Solar System, John Wiley & Sons, New York, p.111.

Prentice A.J.R., 1981, Proc. Astron. Soc. Aust., 4, 164.

Prentice A. J. R., 1984, Earth, Moon Planets, 30, 209.

Prentice A.J.R., 1990, Proc. Astron. Soc. Aust., 8, 364.

Prentice A.J.R., 1991, Proc. Astron. Soc. Aust., 9, 321.

Prentice A.J.R., 1993, Proc. Astron. Soc. Aust., 10, 189.

Prentice A. J. R., 1996a, Earth, Moon Planets, 73, 237.

Prentice A. J. R., 1996b, Eos Trans. AGU, 77 (46), F172.

Prentice, A. J. R., 2001, Earth, Moon Planets, 87, 11.

Prentice A. J. R., 2004a, BAAS, in press.

Prentice A. J. R., 2004b, BAAS, 36, 780.

Prentice A.J.R., Dyt C.P., 2003, MNRAS, 341, 644.

Rappaport N., Bertotti B., Giampieri G., Anderson J.D., 1997, Icarus, 126, 313.

Smagorinsky I., 1963, Mon. Weath. Rev. 91, 99.

Taylor S.R., 1992, Solar System Evolution: A New Perspective, Cambridge Univ. Press, U.K., p.191.

Trujillo, C.A., Brown M.E., 2003, Earth Moon Planets, 92, 99.

Tittemore W. C., 1990, Sci, 250, 263

# TABLES AND TABLE CAPTIONS

**Table 1.** Protosolar cloud mass, initial and present orbital radii and the carbon number distributions of the system of outer protosolar gas rings.

| $n$ | Planet | Elapsed time $t_n$ ($10^5$ yr) | Cloud mass $M_n$ ($M_\odot$) | Initial ring radius $R_{n,i}$ ($R_\odot$) | Present orbit radius $R_n$ ($R_\odot$) | Carbon number fractions $CH_4$ | C(s) | $CO_2$ | CO |
|---|---|---|---|---|---|---|---|---|---|
| 0 | Quaoar | 0.85 | 1.1956 | 7797 | 9323 | 0.2383 | 0.0771 | 0.1974 | 0.4872 |
| 1 | Neptune | 1.41 | 1.1901 | 5431 | 6463 | 0.3692 | 0.0982 | 0.1801 | 0.3525 |
| 2 | Uranus | 1.95 | 1.1832 | 3486 | 4125 | 0.5195 | 0.1120 | 0.1434 | 0.2251 |
| 3 | Saturn | 2.58 | 1.1715 | 1750 | 2050 | 0.8297 | 0.1429 | 0.0132 | 0.0142 |
| 4 | Jupiter | 2.91 | 1.1160 | 964 | 1118 | 0.9597 | 0.0385 | 0.0009 | 0.0009 |

**Table 2.** Physical characteristics of the outer system of proto-solar gas rings and the bulk chemical composition of the condensate.

| Planet | Temperature (K) | Pressure (Pa) | Condensate mass fractions rock | water ice | graphite | $CO_2$ ice | $CH_4$ ice | Condensate mean density (kg m$^{-3}$) |
|---|---|---|---|---|---|---|---|---|
| Quaoar | 26.4 | $1.12 \times 10^{-4}$ | 0.5318 | 0.2056 | 0.0187 | 0.1750 | 0.0689 | 1641 |
| Neptune | 35.8 | $4.68 \times 10^{-4}$ | 0.5414 | 0.2718 | 0.0242 | 0.1626 | 0. | 1834 |
| Uranus | 52.1 | $2.79 \times 10^{-3}$ | 0.5184 | 0.3318 | 0.0264 | 0.1239 | 0. | 1721 |
| Saturn | 94.2 | $4.68 \times 10^{-2}$ | 0.4937 | 0.4742 | 0.0321 | 0. | 0. | 1523 |
| Jupiter | 157.9 | $5.56 \times 10^{-1}$ | 0.5971 | 0.3924 | 0.0105 | 0. | 0. | 1700 |

**Table 3.** Structural and thermal properties of the numerical models of Titan.

| Property | Homogenous model | 2-zone model |
|---|---|---|
| Central pressure (MPa) | 4591 | 6984 |
| Central temperature (K) | 347 | 1520 |
| Central density (kg m$^{-3}$) | 2396 | 3762 |
| Core mass fraction | - | 0.5258 |
| Rock core radius (km) | - | 1678 |
| Core edge temperature (K) | - | 235 |
| Ice convective boundary (km) | 2462 | 2017 |
| Mean density (kg m$^{-3}$) | 2095 | 1904 |
| $C/MR^2$ | 0.383 | 0.316 |

**FIGURES AND FIGURE CAPTIONS**

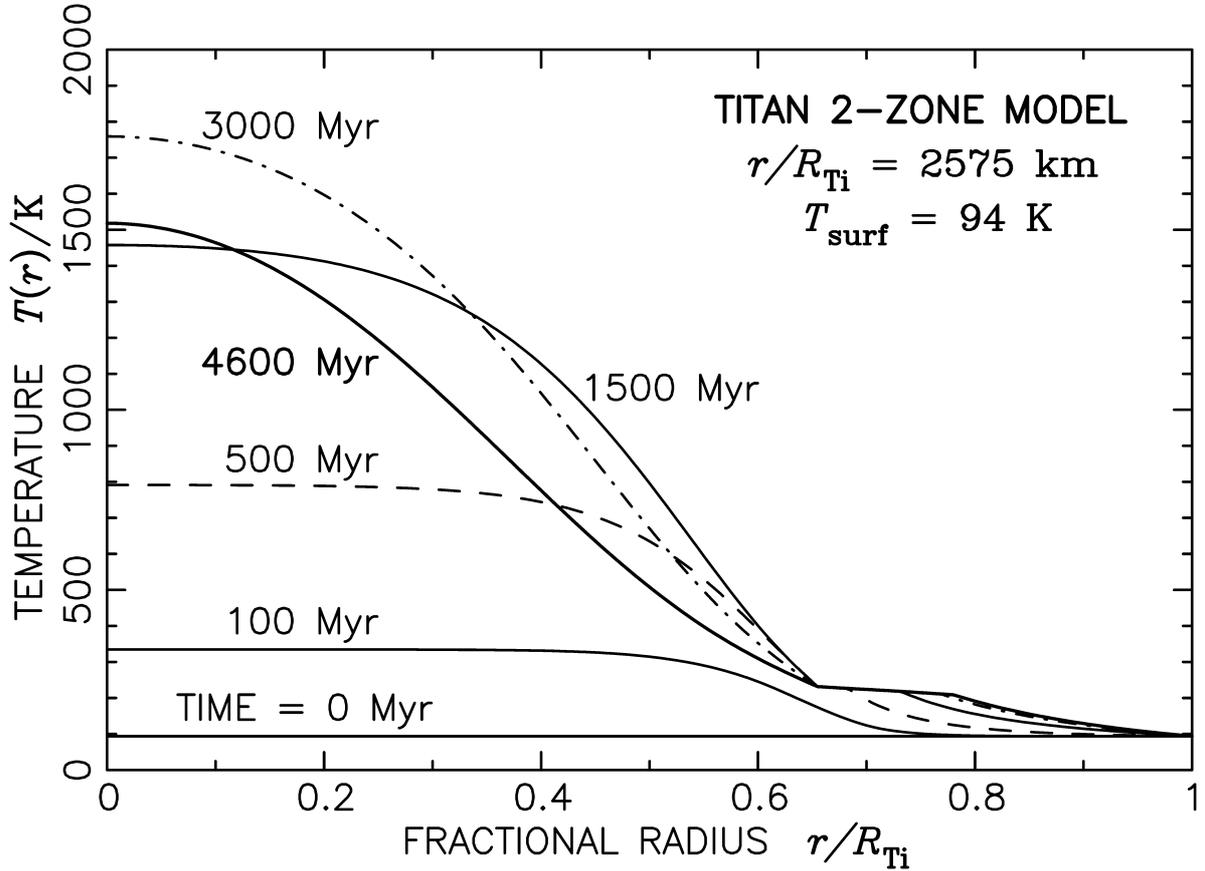

**Figure 1.** Internal temperature profiles of a 2-zone numerical model for Titan versus fractional radius $r/R_{Ti}$ and plotted at several key times during the course of the satellite's thermal evolution to present age (4600 Myr). All heat is derived solely from the decay of radioactive nuclides within the central massive rock-graphite core of mass $0.526 M_{Ti}$. The temperature at the surface $T_{surf}$ is assumed to be constant with time. The $H_2O$ ice mantle becomes thermally convective early on in the evolution. This event first happens at the edge of the central core. Thereafter the ice layer becomes progressively warmer with time but it is never threatened with melting. That is, no internal ocean is predicted in the present day Titan.

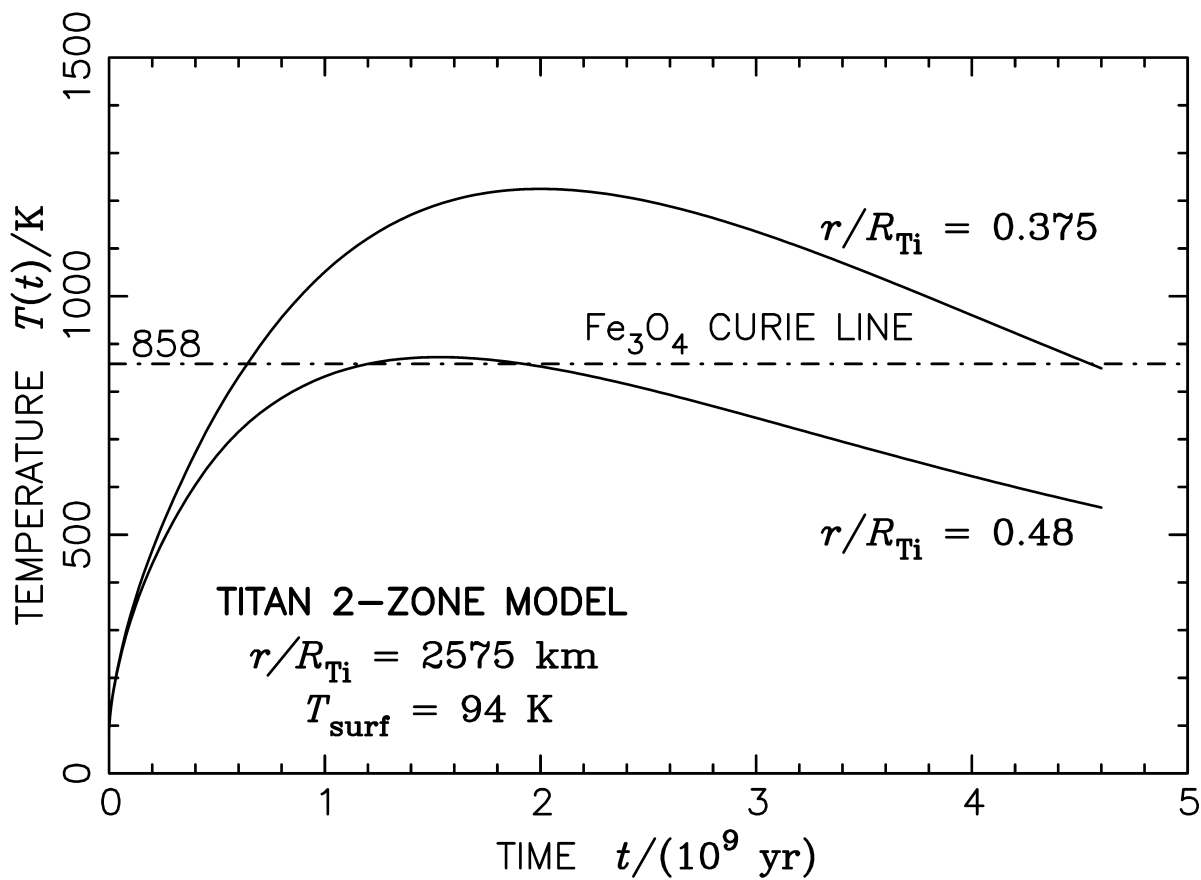

**Figure 2.** The temperature at two internal radius positions within the 2-zone numerical model of Titan plotted as a function of time. Between these two radial positions, the temperature rises above and then descends through the $Fe_3O_4$ Curie point during the course of thermal evolution to present age. It is predicted that Titan may have acquired a permanent internal dipole magnetic field with a moment of magnitude of $\sim 2 \times 10^{11}$ Tesla m$^3$ through thermoremanent magnetization within Saturn's ancient magnetic field.